\documentclass[conference]{IEEEtran}
\usepackage[subpreambles=false]{standalone}
\usepackage{smoothing}
\usepackage{import}
\usepackage{eso-pic}

\AddToShipoutPictureBG*{%
	\put(0,20){
		\hspace*{\dimexpr0.075\paperwidth\relax}%%  change \dimexpr0.5\paperwidth\relax appropriately
		\parbox{.84\paperwidth}{\centering\footnotesize DOI: 10.23919/FUSION52260.2023.10224211 \copyright~2023 IEEE. Personal use of this material is permitted. Permission from IEEE must be obtained for all other uses, in any current or future media, including reprinting/republishing this material for advertising or promotional purposes, creating new collective works, for resale or redistribution to servers or lists, or reuse of any copyrighted component of this work in other works.}%
}}

\IEEEoverridecommandlockouts
\graphicspath{{img/}{sections/img/}}
\begin{document}
	\title{Iterated Filters for Nonlinear Transition Models}
%Dynamically Iterated Filters
	\author{
		\IEEEauthorblockN{Anton Kullberg\IEEEauthorrefmark{1}, Isaac Skog\IEEEauthorrefmark{2},~\IEEEmembership{Senior Member,~IEEE}, and Gustaf Hendeby\IEEEauthorrefmark{1},~\IEEEmembership{Senior Member,~IEEE}.
	\thanks{\noindent This work was partially supported by the Wallenberg AI,
		Autonomous Systems and Software Program (\textsc{WASP}) funded
		by the Knut and Alice Wallenberg Foundation.}
	}
  \IEEEauthorblockA{\IEEEauthorrefmark{1}%
    Dept.~Electrical Engineering, Linköping University, Linköping, Sweden}
	\IEEEauthorblockA{\IEEEauthorrefmark{2}%
		Dept.~Electrical Engineering, Uppsala University, Uppsala, Sweden}
	\IEEEauthorblockA{Email: \ttfamily \{anton.kullberg, gustaf.hendeby\}@liu.se, isaac.skog@angstrom.uu.se}
  }

\maketitle

\begin{abstract}
A new class of iterated linearization-based nonlinear filters, dubbed dynamically iterated filters, is presented.
Contrary to regular iterated filters such as the iterated extended Kalman filter (IEKF), iterated unscented Kalman filter (IUKF) and iterated posterior linearization filter (IPLF), dynamically iterated filters also take nonlinearities in the transition model into account.
The general filtering algorithm is shown to essentially be a (locally over one time step) iterated Rauch-Tung-Striebel smoother.
Three distinct versions of the dynamically iterated filters are especially investigated: analogues to the IEKF, IUKF and IPLF.
The developed algorithms are evaluated on 25 different noise configurations of a tracking problem with a nonlinear transition model and linear measurement model, a scenario where conventional iterated filters are not useful.
Even in this ``simple'' scenario, the dynamically iterated filters are shown to have superior root mean-squared error performance as compared with their respective baselines, the EKF and UKF.
Particularly, even though the EKF diverges in 22 out of 25 configurations, the dynamically iterated EKF remains stable in 20 out of 25 scenarios, only diverging under high noise.
\end{abstract}

% \begin{IEEEkeywords}
% \end{IEEEkeywords}

\section{Introduction}\label{sec:introduction}

State estimation in dynamical systems is a universal problem occurring in the fields of engineering, robotics, economics, etc.
State estimation requires a system model describing the dynamical evolution of the system and a measurement model relating the measured quantities to the state of the system.
If the model is affine with additive Gaussian noise, the most well-known state estimation algorithm is the analytically tractable Kalman filter, which is the optimal estimator in the \emph{mean-squared error} (\abbrMSE) sense \cite{kalmanNewApproachLinear1960}.

In many practical problems, a nonlinear system model is necessary to accurately describe the system.
This means that the state estimation problem is no longer analytically tractable and approximate inference techniques must be used.
Approximate inference in state-space models is a well-studied field in signal processing, machine learning, etc.
Here, we shall focus on linearization-based approximate inference techniques.
These inference techniques linearize the nonlinear model locally (in each time instance) and then employ the Kalman filter.
Analytical linearization leads to the \emph{extended Kalman filter} (\abbrEKF), while sigma-point filters, such as the \emph{unscented Kalman filter} (\abbrUKF) and the \emph{cubature Kalman filter} (\abbrCKF), can be thought of as statistical linearization filters \cite{julierNewApproachFiltering1995, kalmanNewApproachLinear1960, lefebvreCommentNewMethod2002}.

General (Gaussian) state-space models, in the form of a transition model and a measurement model, may equivalently be probabilistically interpreted as a transition density and a measurement density.
Under this interpretation, the linearization-based approximate inference techniques can be thought of as approximating the transition and measurement densities, e.g.,
\begin{equation*}
  \begin{aligned}
    \state_{k+1} &= \dynmod(\state_{k},\pnoise_{k}) &\tikzmarknode{space}{\qquad} p(\state_{k+1}|\state_k) &\overset{a}{\approx} q(\state_{k+1}|\state_k)\\
  \obs_k &= \obsmod(\state_{k}, \onoise_{k}) &\qquad p(\obs_k|\state_k) &\overset{a}{\approx} q(\obs_k|\state_k),
  \end{aligned}
\end{equation*}%
\begin{tikzpicture}[overlay, remember picture]
  \draw[->] (space)++(-0.3,-0.25) to [out=0, in=-180] ++(0.3,0);
\end{tikzpicture}%
where $p(\state_{k+1}|\state_k)$ and $p(\obs_k|\state_k)$ are the transition and measurement density and $q(\state_{k+1}|\state_k)$ and $q(\obs_k|\state_k)$ the corresponding approximations.
Particularly, the linearization-based filters assume affine Gaussian densities for $q(\state_{k+1}|\state_k)$ and $q(\obs_k|\state_k)$ and the Kalman filter is then applied to this ``auxiliary'' model.
The quality of the auxiliary model, and in extension the estimation performance of linearization-based filters, is thus highly dependent on the point (distribution in the statistical case) about which the models are linearized.
Typically, the linearization point (distribution) is chosen to be the mean (distribution) of the current state estimate.
However, a large error in the state estimate can lead to significant linearization errors that may cause even larger estimation errors in the next time step.
This may, in the worst case, cause the filter to diverge.
To alleviate such issues, several variants of iterated filters have been developed, such as the \emph{iterated extended Kalman filter} (\abbrIEKF), the \emph{iterated unscented Kalman filter} (\abbrIUKF) and the \emph{iterated posterior linearization filter} (\abbrIPLF) \cite{jazwinskiStochasticProcessesFiltering1970, skoglundIterativeUnscentedKalman2019, zhanIteratedUnscentedKalman2007, sibleyIteratedSigmaPoint2006, garcia-fernandezPosteriorLinearizationFilter2015}.
These filters essentially iterate the measurement update, where each iteration the measurement model is re-linearized with the ``latest'' iterate.
The research efforts within the field of iterated filters have particularly focused on finding a better linearization point for the measurement model, which is motivated by the fact that nonlinearities in the measurement model (likelihood) affect the resulting state estimate to a greater extent than nonlinearities in the transition model (prior).
Nevertheless, these methods are for instance not useful in the case of a nonlinear transition model but linear measurement model.

In this paper, we seek to fill this gap by developing a class of iterated filters encompassing both the transition model and the measurement model in the iterative process, which we dub dynamically iterated filters.
% These dynamically iterated filters essentially turn out to be different flavors of iterated smoothers applied to the current (and previous) time step.
Note that a dynamically iterated filter based on posterior linearization was first derived in \cite{raitoharjuPosteriorLinearisationFilter2022} for models with non-additive state transition noise. Further, the L-scan \abbrIPLF in \cite{garcia-fernandezIteratedPosteriorLinearization2017} is somewhat similar to the dynamical \abbrIPLF developed here, but requires access to past observations and is thus not strictly a filter.
In this paper, we particularly focus on additive noise models and treat both analytical as well as statistical linearization in a common framework.
The algorithms developed here are essentially dynamically iterated analogues of the \abbrIEKF, \abbrIUKF and \abbrIPLF, as well as other iterated sigma-point filters and does thus not require access to past observations.
These new iterative algorithms encompass both the transition model as well as the measurement model.
Thereby, the proposed algorithms constitute a generalization of conventional iterated filters.
To illustrate the benefits of the proposed algorithms, it is empirically shown that iterating over the transition linearization improves the estimation performance even in the case of a linear measurement model.
Thus, the contributions are twofold:
\begin{itemize}
  \item A detailed derivation of dynamically iterated filters
  \item An extensive numerical evaluation of the developed algorithms as compared to standard nonlinear filters
\end{itemize}

The paper is organized as follows.
In \cref{sec:background}, analytical and statistical linearization as well as the (affine) Kalman smoother equations are restated for completeness.
In \cref{sec:problemformulation}, the state estimation problem is formulated in terms of approximate transition and measurement densities.
\cref{sec:solution} derives the dynamically iterated filters and connects the final solution to iterated (affine) smoothers.
Lastly, \cref{sec:examples} provides a numerical example of the developed algorithm in a tracking scenario where conventional iterated filters are not useful.

% \section{Problem Formulation}\label{sec:problemformulation}
% \import{./sections/}{problemformulation}

\section{Background}\label{sec:background}

For clarity, we here present analytical and statistical linearization in a common framework, as well as restate the well-known Kalman smoother equations.

\subsection{(Affine) Kalman Smoother}
The well-known Kalman filter and \emph{Rauch-Tung-Striebel} (\abbrRTS) smoother equations are repeated here for clarity in terms of a time update, measurement update, and a smoothing step. These can for instance be found in \cite{sarkkaBayesianFilteringSmoothing2013}.
Assume an affine state-space model with additive Gaussian noise, such as
\begin{subequations}
\begin{align}
\state_{k+1} &= \vec{A}_\dynmod\state_k + \vec{b}_\dynmod + \tilde{\pnoise}_k\\
\obs_k &= \vec{A}_\obsmod\state_k + \vec{b}_\obsmod + \tilde{\onoise}_k.
\end{align}
\end{subequations}
Here,  $\tilde{\pnoise}_k\sim\Ndist(\tilde{\pnoise}_k;\vec{0},\vec{Q}\!+\!\boldsymbol{\Omega}_\dynmod)$ and $\tilde{\onoise}_k\sim\Ndist(\tilde{\onoise}_k;\vec{0},\vec{R}\!+\!\boldsymbol{\Omega}_\obsmod)$ are assumed to be mutually independent.
Note that usually, $\boldsymbol{\Omega}_\dynmod=\boldsymbol{\Omega}_\obsmod=\vec{0}$.
For this model, the (affine) Kalman smoother update equations are given by \cref{alg:kalmansmoother}.
\begin{algorithm}[tb]
\caption{(Affine) Kalman smoother}
\label{alg:kalmansmoother}
\begin{enumerate}
\item Time update
\begin{subequations}
\label{eq:timeupdate}
\begin{align}
\hspace*{\dimexpr-\leftmargini}\hat{\state}_{k+1|k} &= \vec{A}_\dynmod\hat{\state}_{k|k} + \vec{b}_\dynmod\\
\hspace*{\dimexpr-\leftmargini}\vec{P}_{k+1|k} &= \vec{A}_\dynmod\vec{P}_{k|k}\vec{A}_\dynmod^\top + \vec{Q} + \boldsymbol{\Omega}_\dynmod.
\end{align}
\end{subequations}
\item Measurement update
\begin{subequations}
\label{eq:measupdate}
\begin{align}
\hspace*{\dimexpr-\leftmargini}\hat{\state}_{k|k} &= \hat{\state}_{k|k-1} + \vec{K}_k(\obs_k - \vec{A}_\obsmod\hat{\state}_{k|k-1} - \vec{b}_\obsmod)\\
\hspace*{\dimexpr-\leftmargini}\vec{P}_{k|k} &= \vec{P}_{k|k-1} - \vec{K}_k\vec{A}_\obsmod\vec{P}_{k|k-1}\\
\hspace*{\dimexpr-\leftmargini}
\vec{K}_k &\triangleq \vec{P}_{k|k-1}\vec{A}_\obsmod^\top(\vec{A}_\obsmod\vec{P}_{k|k-1}\vec{A}_\obsmod^\top + \vec{R} + \boldsymbol{\Omega}_\obsmod)^{-1}.
\end{align}
\end{subequations}
\item Smoothing step
\begin{subequations}
\label{eq:smoothstep}
\begin{flalign}
\hspace*{\dimexpr-\leftmargini}\hat{\state}^s_{k|K} &= \hat{\state}_{k|k} + \vec{G}_k (\hat{\state}^s_{k+1|K} - \hat{\state}_{k+1|k})\\
\hspace*{\dimexpr-\leftmargini}\vec{P}^s_{k|K} &= \vec{P}_{k|k} + \vec{G}_k(\vec{P}^s_{k+1|K} -\nonumber\\ &\quad~ \vec{A}_\dynmod\vec{P}_{k|k}\vec{A}_\dynmod^\top-\vec{Q}-\boldsymbol{\Omega}_\dynmod)\vec{G}_k^\top\\
\hspace*{\dimexpr-\leftmargini}\vec{G}_k &\triangleq  \vec{P}_{k|k}\vec{A}_\dynmod^\top(\vec{A}_\dynmod\vec{P}_{k|k}\vec{A}_\dynmod^\top + \vec{Q} + \boldsymbol{\Omega}_\dynmod)^{-1}
\end{flalign}
\end{subequations}
\end{enumerate}
\end{algorithm}

\subsection{Analytical and Statistical Linearization}
Given a nonlinear model
\begin{equation*}
\vec{z} = \vec{g}(\state),
\end{equation*}
we wish to find an affine representation
\begin{equation}
\vec{g}(\state)\approx \vec{A}\state + \vec{b} + \eta,
\end{equation}
with $\eta\sim\Ndist(\eta; \vec{0}, \boldsymbol{\Omega})$.
In this affine representation, there are three free parameters: $\vec{A}, \vec{b}$ and $\boldsymbol{\Omega}$.
Analytical linearization through first-order Taylor expansion selects the parameters as
\begin{equation}
\vec{A}=\frac{d}{d\state}\vec{g}(\state)|_{\state=\bar{\state}},
\quad \vec{b} = \vec{g}(\state)|_{\state=\bar{\state}} - \vec{A}\bar{\state},
\quad \boldsymbol{\Omega}=\vec{0},
\end{equation}
where $\bar{\state}$ is the point about which the function $\vec{g}(\state)$ is linearized.
Note that $\boldsymbol{\Omega}=\vec{0}$ essentially implies that the linearization is assumed to be error free.

Statistical linearization instead linearizes w.r.t. a distribution $p(\state)$.
Assuming that such a distribution $p(\state)=\Ndist(\state;\hat{\state},\vec{P})$ is given, statistical linearization selects the affine parameters as
\begin{subequations}
\begin{align}
\vec{A} &= \Psi^\top \vec{P}^{-1}\\
\vec{b} &= \bar{\vec{z}}-\vec{A}\hat{\state}\\
\boldsymbol{\Omega} &= \Phi-\vec{A} \vec{P} \vec{A}^\top\\
\bar{\vec{z}} &= \mathbb{E}[\vec{g}(\state)]\\
\Psi &= \mathbb{E}[(\state-\hat{\state})(\vec{g}(\state) - \bar{\vec{z}})^\top]\\
\Phi &= \mathbb{E}[(\vec{g}(\state) - \bar{\vec{z}})(\vec{g}(\state) - \bar{\vec{z}})^\top],
\end{align}
\end{subequations}
where the expectations are taken w.r.t. $p(\state)$.
The major difference from analytical linearization is that ${\boldsymbol{\Omega}\neq 0}$, implying that the error in the linearization is captured.

\section{Problem Formulation}\label{sec:problemformulation}

To set the stage for the algorithm
development, the general state estimation problem is described here with a probabilistic viewpoint. To that end, consider a discrete-time state-space model (omitting a possible input $\inp_k$ for notational brevity) given by
\begin{subequations}\label{eq:ssm}
\begin{align}
\state_{k+1} &= \dynmod(\state_{k}) + \pnoise_{k}\label{eq:dyneq}\\
\obs_k &= \obsmod(\state_k) + \onoise_k\label{eq:obseq}\\
p(\pnoise_k) &= \Ndist(\pnoise_k;\vec{0}, \vec{Q}),\quad
p(\onoise_k) = \Ndist(\onoise_k;\vec{0}, \vec{R}).
\end{align}
\end{subequations}
Here, $\state_k,~\obs_k,~\pnoise_k$ and $\onoise_k$ denote the state, the measurement, the process noise and the measurement noise at time $k$, respectively.
It is further assumed that $\state_k\in \mathcal{X},\forall k$ and that $\pnoise_k$ and $\onoise_k$ are mutually independent.
Note that \cref{eq:dyneq,eq:obseq} can equivalently be written as a \emph{transition density} and a \emph{measurement density} as
\begin{subequations}
\begin{align}
p(\state_{k+1}|\state_{k}) &= \Ndist(\state_{k+1};\dynmod(\state_{k}), \vec{Q})\label{eq:transitiondensity}\\
p(\obs_k|\state_k) &= \Ndist(\obs_k;\obsmod(\state_k), \vec{R})\label{eq:observationdensity}.
\end{align}
\end{subequations}
Further, the initial state distribution is assumed to be given by
\begin{equation}
p(\state_0)=\Ndist(\state_0;\hat{\state}_{0|0}, \vec{P}_{0|0}).
\end{equation}

Given the transition and measurement densities and a sequence of measurements $\obs_{1:k}=\begin{bmatrix}\obs_1^\top&\dots&\obs_k^\top\end{bmatrix}^\top$, the state estimation problem consists of computing the posterior of the state sequence (trajectory), i.e., computing
\begin{equation}\label{eq:smoothingproblem}
p(\state_{0:k}|\obs_{1:k}) = \frac{1}{\vec{Z}_{1:k}} p(\state_0)\prod_{i=1}^{k} p(\obs_i|\state_i)p(\state_i|\state_{i-1}),
\end{equation}
where
\begin{equation*}
\vec{Z}_{1:k}=\int_{\mathcal{X}} p(\state_0)\prod_{i=1}^{k} p(\obs_i|\state_i)p(\state_i|\state_{i-1}) d\state_0 \cdots d\state_k,
\end{equation*}
is the marginal likelihood of $\obs_{1:k}$.
The posterior \eqref{eq:smoothingproblem} is commonly referred to as the joint \emph{smoothing} distribution which, in the case of linear $\dynmod$ and $\obsmod$, can be analytically found through the Kalman smoother, e.g., the \abbrRTS smoother \cite{sarkkaBayesianFilteringSmoothing2013}.

In the setting considered here, i.e., in \emph{filtering} applications, the densities of interest are rather the \emph{marginal} posteriors
\begin{equation}\label{eq:statemarginaltimek}
p(\state_k|\obs_{1:k}) = \frac{p(\obs_k|\state_k)\!\int_{\mathcal{X}}\! p(\state_k|\state_{k-1})p(\state_{k-1}|\obs_{1:k-1})d\state_{k-1}}{\vec{Z}_{k}},
\end{equation}
for all times $k$, where
\begin{equation*}
\vec{Z}_{k} = \int_{\mathcal{X}} p(\obs_k|\state_k) p(\state_k|\state_{k-1})p(\state_{k-1}|\obs_{1:k-1})d\state_{k-1}d\state_k.
\end{equation*}
Again, in the case of linear $\dynmod$ and $\obsmod$, the (analytical) solution is given by the Kalman filter \cite{kalmanNewApproachLinear1960}.

In the general case, the marginal posteriors can not be computed analytically.
Inspecting \cref{eq:statemarginaltimek}, there are two integrals that require attention.
We turn first to the Chapman-Kolmogorov equation
\begin{equation}\label{eq:chapman}
p(\state_k|\obs_{1:k-1}) = \int_{\mathcal{X}} p(\state_k|\state_{k-1})p(\state_{k-1}|\obs_{1:k-1})d\state_{k-1}.
\end{equation}
Assuming that $p(\state_{k-1}|\obs_{1:k-1})$ is Gaussian, \cref{eq:chapman} has a closed form solution given by \cref{eq:timeupdate}, \emph{if}
$p(\state_k|\state_{k-1})$ is Gaussian and \cref{eq:dyneq} is affine.
Therefore, as \cref{eq:transitiondensity} is Gaussian, we seek an affine approximation of the transition function $\dynmod$ as
\begin{equation}\label{eq:dynapprox}
\dynmod(\state_{k-1}) \approx \vec{A}_\dynmod\state_{k-1} + \vec{b}_\dynmod + \eta_\dynmod,
\end{equation}
with $p(\eta_\dynmod) = \Ndist(\eta_\dynmod;\vec{0},\boldsymbol{\Omega}_\dynmod)$.
Hence, the transition density $p(\state_{k}|\state_{k-1})$ is approximated by $q(\state_{k}|\state_{k-1})$ as
\begin{equation}\label{eq:transdensityapprox}
q(\state_{k}|\state_{k-1}) = \Ndist(\state_{k}; \vec{A}_\dynmod\state_{k-1}+\vec{b}_\dynmod, \vec{Q}+\boldsymbol{\Omega}_\dynmod).
\end{equation}
If $\vec{A}_\dynmod,\vec{b}_\dynmod$ and $\boldsymbol{\Omega}_\dynmod$ are chosen to be the analytical linearization of $\dynmod$ around the mean of the posterior $p(\state_{k-1}|\obs_{1:k-1})$, the \abbrEKF time update is recovered through \cref{eq:timeupdate}.
Similarly, statistical linearization around the posterior at time $k-1$ recovers the sigma-point filter time updates.
This yields an approximate predictive distribution $q(\state_k|\obs_{1:k-1})$, which can then be used to approximate the second integral of interest (and subsequently, the posterior at time $k$).
Explicitly, the second integral is given by
\begin{equation}\label{eq:normalizationapprox}
\vec{Z}_{k} \approx \int_{\mathcal{X}} p(\obs_k|\state_k)q(\state_k|\obs_{1:k-1})d\state_k.
\end{equation}
Similarly to \cref{eq:dynapprox}, \cref{eq:normalizationapprox} has a closed form solution if $p(\obs_k|\state_k)$ is Gaussian and \cref{eq:obseq} is affine.
Thus, as \cref{eq:observationdensity} is Gaussian, we seek an affine approximation of the measurement function $\obsmod$ as
\begin{equation}\label{eq:obsapprox}
\obsmod(\state_k) \approx \vec{A}_\obsmod\state_k + \vec{b}_\obsmod + \eta_\obsmod,
\end{equation}
with $p(\eta_\obsmod) = \Ndist(\eta_\obsmod; \vec{0}, \boldsymbol{\Omega}_\obsmod)$.
Hence, the measurement density $p(\obs_k|\state_k)$ is approximated by $q(\obs_k|\state_k)$ as 
\begin{equation}\label{eq:obsdensityapprox}
q(\obs_k|\state_k) = \Ndist(\obs_k; \vec{A}_\obsmod\state_k + \vec{b}_\obsmod, \vec{R} + \boldsymbol{\Omega}_\obsmod),
\end{equation}
which leads to an analytically tractable integral.
With \cref{eq:transdensityapprox,eq:obsdensityapprox}, the (approximate) marginal posterior \cref{eq:statemarginaltimek} is now given by
\begin{equation}\label{eq:statemarginaltimekapprox}
q(\state_k|\obs_{1:k}) = \frac{q(\obs_k|\state_k)q(\state_k|\obs_{1:k-1})}{\int_{\mathcal{X}} q(\obs_k|\state_k)q(\state_k|\obs_{1:k-1})d\state_k},
\end{equation}
which is analytically tractable and given by \cref{eq:measupdate}.
Note that analytical linearization of \cref{eq:obsapprox} around the mean of $q(\state_{k}|\obs_{1:k-1})$ recovers the \abbrEKF measurement update whereas statistical linearization recovers the sigma-point measurement update(s).

The quality of the approximate marginal posterior \cref{eq:statemarginaltimekapprox} thus directly depends on the quality of the approximations \cref{eq:transdensityapprox,eq:obsdensityapprox}.
The quality of \cref{eq:transdensityapprox,eq:obsdensityapprox} in turn directly depends on the choice of linearization points or densities, which is typically chosen to be the approximate predictive and previous approximate posterior distributions.
This choice is of course free and iterative filters such as the \abbrIEKF, \abbrIUKF or \abbrIPLF have been proposed to improve these approximations \cite{jazwinskiStochasticProcessesFiltering1970, garcia-fernandezIteratedStatisticalLinear2014, skoglundIterativeUnscentedKalman2019, zhanIteratedUnscentedKalman2007}.
These filters can be thought of as finding an approximate posterior $q^i(\state_k|\obs_{1:k})$ which is then used to re-linearize the function $\obsmod$, producing a new approximation $q^{i+1}(\state_k|\obs_{1:k})$.
This is then iterated until some convergence criterion is satisfied; typically until a fixed point is reached or a maximum number of iterations has been reached.

However, none of these algorithms, except \cite{raitoharjuPosteriorLinearisationFilter2022}, encompass the approximate density \cref{eq:transdensityapprox}, even though this approximation directly affects the approximate marginal posterior as well.
This is motivated by the fact that nonlinearities in the likelihood affect the posterior approximation to a greater extent than the prior.
Nevertheless, standard iterated filters are for instance not useful in the case of a nonlinear transition function $\dynmod$ but linear measurement function $\obsmod$, even though the linearization of $\dynmod$ also affects the quality of the approximate posterior.
Next, a general linearization-based algorithm encompassing both the transition density as well as the measurement density approximations is developed.

\section{Dynamically Iterated Filter}\label{sec:solution}

To derive an algorithm encompassing both the transition density \cref{eq:transdensityapprox}, as well as the observation density \cref{eq:obsdensityapprox}, at time $k$, we naturally need to seek an approximate posterior over both $\state_{k-1}$ as well as $\state_k$.
To do so, we generalize the derivation in \cite{garcia-fernandezPosteriorLinearizationFilter2015} to extend backwards one step.
Define two auxiliary variables, $\auxil_k,~\auxil_{k-1}$ as
\begin{subequations}
\begin{align}
\auxil_{k-1} &= \dynmod(\state_{k-1}) + \psi\\
\auxil_k &= \obsmod(\state_k) + \phi\\
p(\psi) &= \Ndist(\vec{0}, \alpha\vec{I}),\quad
p(\phi) = \Ndist(\vec{0}, \beta\vec{I}),
\end{align}
\end{subequations}
where $\psi$ and $\phi$ are independent of each other as well as the process noise $\pnoise$ and the measurement noise $\onoise$.
Note that as $\alpha,~\beta\to 0$, $\auxil_{k-1}\to \dynmod(\state_{k-1})$ and $\auxil_k\to\obsmod(\state_k)$.
Now, the true joint posterior of $\state_{k-1},~\state_k,~\auxil_{k-1}$ and $\auxil_k$ is given by
\begin{align}\label{eq:jointdecomposition}
&\fulljoint \propto\nonumber \\&\fulljointexpand.
\end{align}
Following \cite{garcia-fernandezPosteriorLinearizationFilter2015}, we assume that the approximate posterior can be decomposed in the same manner, i.e.,
\begin{align}\label{eq:approximatejoint}
 &\approxjoint \approx \nonumber\\&\approxjointexpand,
\end{align}
where $\theta$ are the parameters of the affine approximation of the transition model and measurement model, i.e., ${\theta=[\vec{A}_\dynmod,\vec{b}_\dynmod,\boldsymbol{\Omega}_\dynmod, \vec{A}_\obsmod,\vec{b}_\obsmod,\boldsymbol{\Omega}_\obsmod]}$.

We now seek a $\theta$ such that $\approxjoint$ is close to $\fulljoint$, in some sense.
Formally, the optimal parameters $\theta^*$, and hence the optimal affine approximations of $\dynmod$ and $\obsmod$, are found through
\begin{align}\label{eq:fullopt}
\theta^* = & \argmin_\theta\Loss.
\end{align}
The loss $\Loss$ is free to choose, but a natural choice of dissimilarity measure between distributions is the Kullback--Leibler (\abbrKL) divergence, which we pursue here.
The \abbrKL divergence between the true joint posterior and the approximate joint posterior is given by
\begin{multline}
\mathrm{KL}\left(\fulljoint \Vert \approxjoint \right)=\\
\mathrm{KL}\left( \statejoint \Vert \approxstatejoint \right) +\\
\mathbb{E}\left[\mathrm{KL}(\auxilcond{k}\Vert\approxauxilcond{k})\right] +\\
\mathbb{E}\left[
\mathrm{KL}(\auxilcond{k-1}\Vert\approxauxilcond{k-1})
\right] \triangleq\Loss. \label{eq:fulloss}
\end{multline}
See \cref{app:loss} for the derivation.
Note that the expectations in $\Loss$ are taken with respect to the true joint posterior $\statejoint$.
It is noteworthy that $\Loss$ can be decomposed into three distinct terms, each dealing with each respective factor of \cref{eq:approximatejoint}.
The first term is simply the \abbrKL divergence between the true and approximate joint posterior of the states at time $k$ and $k-1$.
The second and third terms are the expected \abbrKL divergences of the affine approximation of the measurement model and transition model, respectively, where the expectation is taken with respect to the true joint posterior $\statejoint$.

It is impractical to minimize \cref{eq:fulloss}, seeing as the expectations are taken w.r.t. the true joint posterior $\statejoint$.
Nevertheless, an iterative procedure may be used to approximately solve this minimization problem.
% Seeing as the first term of $\Loss$ is $\geq 0$, a lower bound is given by the second term, i.e.,
% \begin{equation*}
% \Loss \geq \mathbb{E}
% \left[
% \mathrm{KL}(\auxilcond{k}\Vert\approxauxilcond{k}) + \mathrm{KL}(\auxilcond{k-1}\Vert\approxauxilcond{k-1})
% \right].
% \end{equation*}
% Hence, the optimal linear approximations of $f$ and $h$ are given by
% \begin{equation*}
% \theta^* = \arg\min_\theta \mathbb{E}_{\statejoint}
% \left[
% \mathrm{KL}(\auxilcond{k}\Vert\approxauxilcond{k}) + \mathrm{KL}(\auxilcond{k-1}\Vert\approxauxilcond{k-1})
% \right],
% \end{equation*}
% where $\theta$ includes all the parameters of the linearization(s).

% Schematic illustration of a dynamically iterated filter
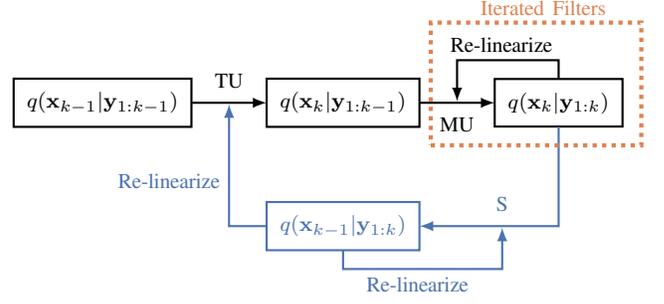
\begin{figure}[tb]
\centering
\begin{tikzpicture}[squarednode/.style={rectangle, draw=black, text opacity=1, thick, minimum size=5mm},
every node/.style={inner sep=5pt,outer sep=0pt, font=\footnotesize}]
\node[squarednode] (previous) {$q(\state_{k-1}|\obs_{1:k-1})$};
\node[squarednode] (predicted) [right=of previous] {$q(\state_{k}|\obs_{1:k-1})$};
\node[squarednode] (updated) [right=of predicted] {$q(\state_{k}|\obs_{1:k})$};
\node[squarednode, draw=snsblue, text=snsblue] (smoothed) [below=of predicted] {$q(\state_{k-1}|\obs_{1:k})$};
% Prediction step
\draw[-latex, thick] (previous.east) -- node[inner sep=0pt, label=above:{TU}] (timeupdate) {} (predicted.west);
% Update step
\draw[-latex, thick] (predicted.east) -- node[label=below:{MU}, inner sep=0pt] (measurementupdate) {} (updated.west);
% Smoothing step
\draw[-latex, thick, snsblue] (updated.south) |-  node[inner sep=0pt, xshift=-.75cm, label=above:{S}] (smoothstep) {} (smoothed.east);
% Prediction re-linearization
\draw[-latex, thick, snsblue] (smoothed.west) -| node[label=left:{Re-linearize}, yshift=.6cm, xshift=.25cm] {} (timeupdate.south);
% Update re-linearization
\draw[-latex, thick] (updated.north) |- ++(0, 0.25) -| node[label=above:{Re-linearize}, xshift=.6cm, yshift=-.25cm] {} (measurementupdate.north);
% Smoothing re-linearization
\draw[-latex, thick, snsblue] (smoothed.south) |- ++(0, -0.25) node[xshift=1cm, yshift=.25cm, label=below:{Re-linearize}] {} -| (smoothstep.south);
% Rectangle around update step
\draw[ultra thick, dotted, snsorange] (updated.north west)+(-.85, .75)
node [xshift=1.5cm, yshift=-.25cm, label=above:{Iterated Filters}] {}
rectangle ($(updated.south east)+(0.25, -0.25)$);
\end{tikzpicture}
\caption{Schematic illustration of a dynamically iterated filter. Ordinary iterated filters, marked in dotted orange, only re-linearize the measurement update. Dynamically iterated filters also re-linearize the time update through a smoothing step, marked in blue. The time update (TU) and the smoothing step (S) are linearized w.r.t. the smoothed distribution $q(\state_{k-1}|\obs_{1:k})$. The measurement update (MU) is linearized w.r.t. the current posterior $q(\state_k|\obs_{1:k})$. The steps are iterated until some convergence criterion is met.}
\label{fig:dyniterfilter}
\end{figure}

\subsection{Iterative Solution}
To practically optimize \cref{eq:fullopt}, we assume access to an $i$:th approximation to the state joint posterior $\statejoint\approx\approxstatejoint[i]$.
We then use $\approxstatejoint[i]$ in place of $\statejoint$ in \cref{eq:fulloss} and thus optimize an approximate loss, i.e.,
the approximate optimization problem is given by
\begin{multline*}
\theta^* =  \argmin_\theta
\mathrm{KL}\left( \approxstatejoint[i] \Vert \approxstatejoint[i+1] \right) +\\
\mathbb{E}_{\approxstatejoint[i]}\left[\mathrm{KL}(\auxilcond{k}\Vert\approxauxilcond[i+1]{k})\right] +\\
\mathbb{E}_{\approxstatejoint[i]}
\left[
\mathrm{KL}(\auxilcond{k-1}\Vert\approxauxilcond[i+1]{k-1})
\right],
\end{multline*}
where the expectations are now over $\approxstatejoint[i]$.
Sufficiently close to a fixed point, the first \abbrKL term is approximately 0 and the final optimization problem is thus given by
\begin{multline}
\theta^* = \argmin_\theta
\mathbb{E}_{\approxstatejoint[i]}\biggl[\mathrm{KL}(\auxilcond{k}\Vert\approxauxilcond[i+1]{k}) \\
+\mathrm{KL}(\auxilcond{k-1}\Vert\approxauxilcond[i+1]{k-1})
\biggr].
\end{multline}
Technically, the optimal $\theta^*$ is given by statistical linearization of $\dynmod$ and $\obsmod$ w.r.t. the current approximation $\approxstatejoint[i]$, see e.g., \cite{garcia-fernandezPosteriorLinearizationFilter2015}.
Note that statistical linearization of $\dynmod$ w.r.t. $\approxstatejoint[i]$ only requires the marginal $\approxstate{i}{k-1}$.
Similarly, statistical linearization of $\obsmod$ only requires the marginal $\approxstate{i}{k}$.
Thus, the algorithm conceptually amounts to predicting forward in time, performing a measurement update and smoothing backwards in time in order to provide new linearization points (densities) for both the transition density as well as the measurement density simultaneously.
These steps are then iterated until fixed point convergence, finally providing an approximate posterior $q(\state_{k-1:k}|\obs_{1:k})$.
The general algorithm is summarized in \cref{alg:dyniterfilter} and schematically depicted in \cref{fig:dyniterfilter}.

% An algorithmic description of a dynamically iterated filter
\begin{algorithm}
\caption{Dynamically iterated filter}
\label{alg:dyniterfilter}
\begin{algorithmic}
\Require $q(\state_{k-1}|\obs_{1:k-1})=\Ndist(\state_{k-1};\hat{\state}_{k-1|k-1},\vec{P}_{k-1|k-1})$
\State Compute $\hat{\state}_{k|k-1}^0, \vec{P}_{k|k-1}^0$ by \cref{eq:timeupdate}
\State Compute $\hat{\state}_{k|k}^0, \vec{P}_{k|k}^0$ by \cref{eq:measupdate}
\State Compute $\hat{\state}_{k-1|k}^0, \vec{P}_{k-1|k}^0$ by \cref{eq:smoothstep}
\State $i\gets 0$
\While{not converged}
  \State Calculate $(\vec{A}_\dynmod,\vec{b}_\dynmod, \boldsymbol{\Omega}_\dynmod)$ by analytical or statistical linearization of $\dynmod$ about $\hat{\state}_{k-1|k}^i, \vec{P}_{k-1|k}^i$
  \State Compute $\hat{\state}_{k|k-1}^{i+1}, \vec{P}_{k|k-1}^{i+1}$ by \cref{eq:timeupdate}
  \State Calculate $(\vec{A}_\obsmod,\vec{b}_\obsmod, \boldsymbol{\Omega}_\obsmod)$ by analytical or statistical linearization of $\obsmod$ about $\hat{\state}_{k|k}^i, \vec{P}_{k|k}^i$
  \State Compute $\hat{\state}_{k|k}^{i+1}, \vec{P}_{k|k}^{i+1}$ by \cref{eq:measupdate}
  \State Compute $\hat{\state}_{k-1|k}^{i+1}, \vec{P}_{k-1|k}^{i+1}$ by \cref{eq:smoothstep}
  \State $i\gets i+1$
\EndWhile
\State \textbf{return} $\hat{\state}_{k|k}^{i}, \vec{P}_{k|k}^{i}, \hat{\state}_{k-1|k}^{i}, \vec{P}_{k-1|k}^{i}$
\end{algorithmic}
\end{algorithm}

Note that the algorithm is applicable to all possible combinations of models with linear and nonlinear $\dynmod$ and $\obsmod$.
Further, even though the developed solution is essentially an \abbrIPLF also encompassing the transition density, by changing the linearization method from statistical to analytical, an ``extended'' version is recovered in similar spirit to the \abbrIEKF.
Furthermore, an \abbrIUKF version, similar to \cite{skoglundIterativeUnscentedKalman2019}, may also be recovered by ``freezing'' the covariance matrices $\vec{P}_{k-1|k}^i=\vec{P}_{k-1|k-1}$ and $\vec{P}_{k|k}^i=\vec{P}^i_{k|k-1}$ and only updating these during the last iteration.
It is also worthwhile to point out that the dynamically iterated filters are essentially ``local'' iterated smoothers, analogous to the \emph{iterated extended Kalman smoother} (\abbrIEKS) \cite{bellIteratedKalmanSmoother1994a} and the \emph{iterated posterior linearization smoother} (\abbrIPLS) \cite{garcia-fernandezIteratedPosteriorLinearization2017}, operating on just one time instance and observation.
Therefore, as noted in \cite{raitoharjuPosteriorLinearisationFilter2022}, a byproduct of the algorithm is a one-step smoothed state estimate and the method can thus be thought of as an iterated one-step fixed-lag smoother as well.

All that is left is to determine a stopping criterion for the iterations.
Similarly to \cite{garcia-fernandezPosteriorLinearizationFilter2015}, a stopping criterion for the iterative updates may be formed on the basis of the \abbrKL divergence between two successive approximations of the posterior, i.e.,
\begin{equation*}
\mathrm{KL}(q^i(\state_k|\obs_{1:k})\Vert q^{i+1}(\state_k|\obs_{1:k})) < \gamma.
\end{equation*}
Another possibility to check for fixed-point convergence is to instead use the smoothed density $q(\state_{k-1}|\obs_{1:k})$ in a similar manner as the posterior.
This is not investigated in detail here.
Instead, in the numerical example in \cref{sec:examples}, a fixed number of iterations are used for simplicity.

\section{Numerical Examples}\label{sec:examples}

To demonstrate the application of the dynamically iterated filters, we provide an illustrative example demonstrating the iterative procedure of the algorithm.
We also provide a numerical example of maneuvering target tracking with a nonlinear transition model but a \emph{linear} measurement model.

\subsection{Illustrative example}
\begin{figure}[tb]
\centering\hspace{-.5cm}
\begin{tikzpicture}[every node/.style={inner sep=0pt,outer sep=0pt, font=\footnotesize}]
\node[label=above:{Smoothed / Prior}] (smooth) {\includegraphics[width=.9\columnwidth]{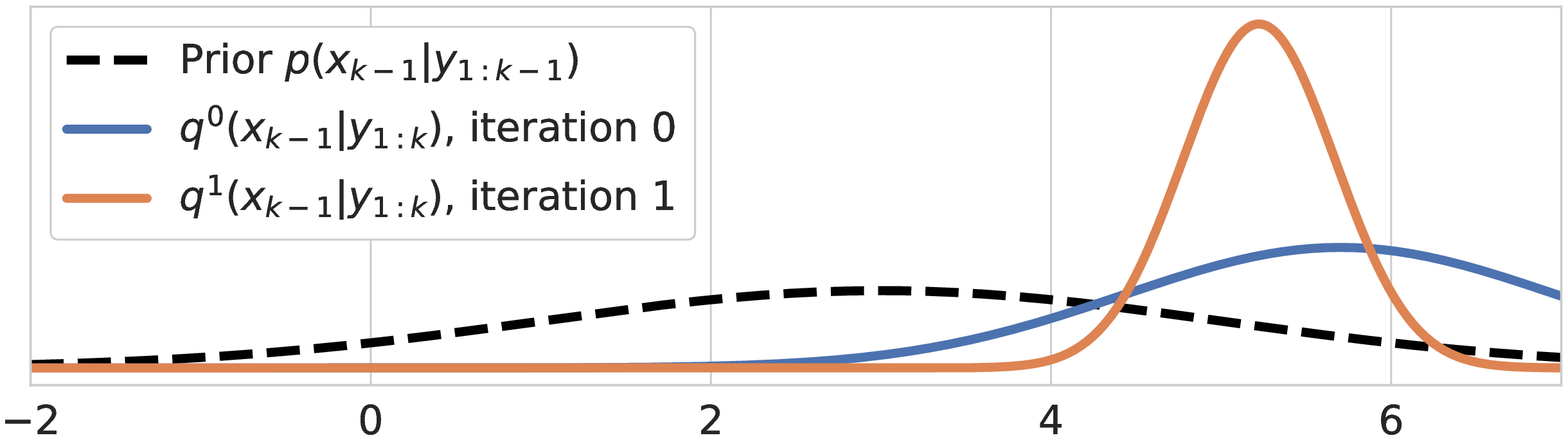}};
\node[below= of smooth, label=above:{Predictive}, yshift=.75cm] (pred) {\includegraphics[width=.9\columnwidth]{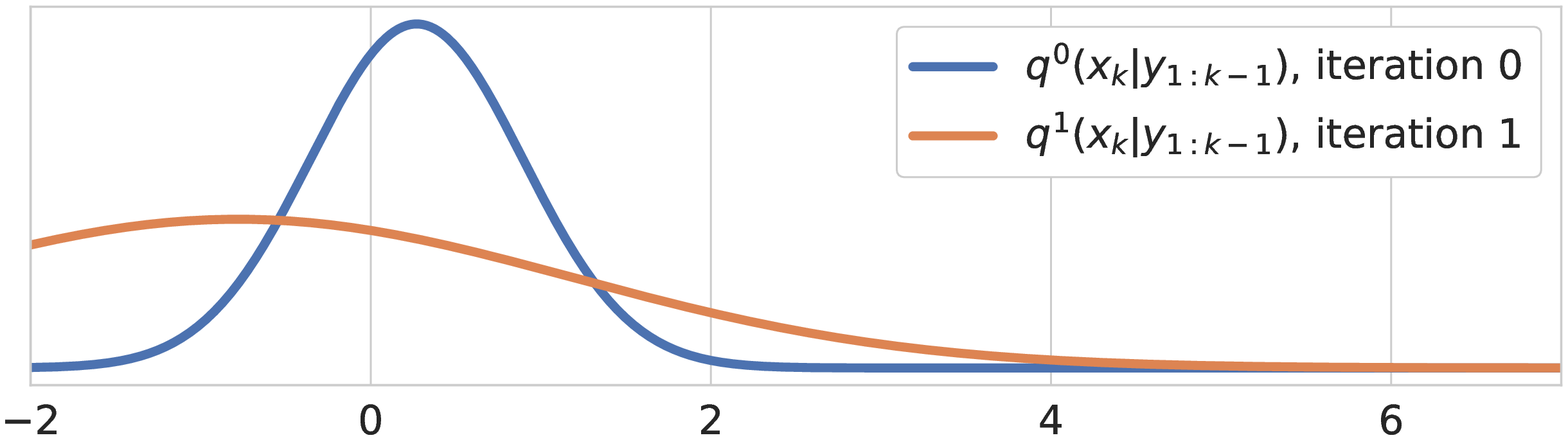}};
\node[below= of pred, label=above:{Posterior}, yshift=.75cm] (post) {\includegraphics[width=.9\columnwidth]{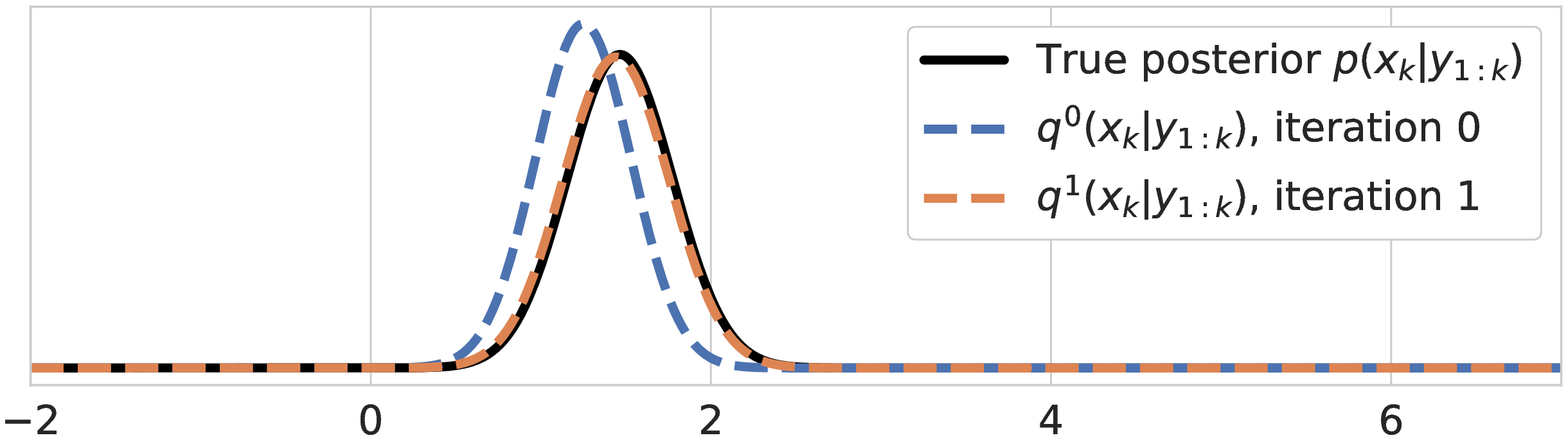}};
% \draw[-latex, thick] (smooth.east) to [out=-65, in=65] (pred.east) node[left] {TU};
\draw[-, thick] (smooth.east)++(0,-0.25) edge[-latex, out=-65, in=65] node[left,xshift=-.25cm, yshift=.25cm] {TU} (pred.east);
% \draw[-latex, thick] (pred.east)++(0,-0.25) to [out=-65, in=65] (post.east);
\draw[-, thick] (pred.east)++(0,-0.25) edge[-latex, out=-65, in=65] node[left, xshift=-.25cm, yshift=.25cm] {MU} (post.east);
% \draw[-latex, thick] (post.west) to [out=105, in=-105] (smooth.west);
\draw[-, thick] (post.west) edge[-latex, out=105, in=-105] node[right, xshift=.25cm] {S} (smooth.west);
\end{tikzpicture}
\caption{Illustration of the (extended) dynamically iterated filter.
The black curves in the top and bottom plots are the prior $p(\state_{k-1}|\obs_{1:k-1})$ and true posterior $p(\state_{k}|\obs_{1:k})$, respectively.
The blue curves from top to bottom illustrate the approximate smoothed, predictive and posterior densities at iteration 0, respectively.
The orange curves illustrate the same densities during the second iteration of the filter.
The filter thus moves from the prior (top) to the predictive (middle) to the posterior (bottom) and back up to the smoothed (top).
The time update, measurement update and smoothing step are indicated similarly to \cref{fig:dyniterfilter}.
Notice that iteration 0 exactly corresponds to an \abbrEKF.}
\label{fig:illustrativeexample}
\end{figure}

To illustrate the iterative procedure of the algorithm, we use an example similar to that in \cite{garcia-fernandezIteratedStatisticalLinear2014} but alter it to include a dynamical model.
Therefore, let the model be given by
\begin{align*}
\state_{k+1} &\sim \Ndist(\state_{k+1}; a\state_k^3, Q)\\
\obs_k &\sim \Ndist(\obs_k; \state_k, R),
\end{align*}
with $a=0.01, Q=0.1$ and $R=0.1$.
We assume that a prior is given at time $k-1$ as $p(\state_{k-1}|\obs_{1:k-1})=\Ndist(\state_{k-1}; 3, 4)$.
We then apply an analytically linearized version of the dynamically iterated filter to this model and plot the intermediary and final approximate predictive, posterior, and smoothed densities. The true posterior is found simply through evaluating the posterior density over a dense grid.
The example is illustrated in \cref{fig:illustrativeexample}, where two iterations are enough for the posterior approximation to be accurate.

\subsection{Maneuvering Target Tracking}
We consider a numerical example of maneuvering target tracking with a nonlinear transition model but a \emph{linear} measurement model.
This is a typically ``easy'' tracking scenario where standard filters generally do well.

Three versions of the dynamically iterated filters are evaluated, an extended version (\abbrDIEKF), an unscented version (\abbrDIUKF), and a posterior linearization version (\abbrDIPLF) based on unscented transform.
These are compared to their respective non-iterated counterparts, i.e., the \abbrEKF and the \abbrUKF.
For the unscented filters, we use the tuning parameters $\alpha=\sqrt{3/n_x}, \kappa=\frac{n_x(3/2-\alpha^2)}{\alpha^2}$ and $\beta=2$, where $n_x$ is the dimension of $\state$.
This tuning corresponds to a weighting of $1/3$ on the central sigma point.

We consider a target maneuvering in a plane and describe the target using the state vector $\state_k^\top=\begin{bmatrix}p^x_k & v^x_k & p^y_k & v^y_k & \omega_k \end{bmatrix}$. Here, $p^x_k,~p^y_k,~v^x_k,~v^y_k$ are the Cartesian coordinates and velocities of the target, respectively, and $\omega_k$ is the turn rate at time $k$.
The transition model is thus given by
\begin{equation}
\state_{k+1} = \vec{F}(\omega_k)\state_k + \pnoise_k,
\end{equation}
where
\begin{equation*}
\vec{F}(\omega_k) =
\begin{bmatrix}
1 & \frac{\sin(T \omega_k)}{\omega_k} & 0 & -\frac{(1-\cos(T \omega_k ))}{\omega_k} & 0\\
0 & \cos(T \omega_k ) & 0 & -\sin(T \omega_k ) & 0\\
0 & \frac{(1-\cos(T \omega_k ))}{\omega_k} & 1 & \frac{\sin(T \omega_k )}{\omega_k} & 0\\
0 & \sin(T \omega_k ) & 0 & \cos(T \omega_k ) & 0\\
0 & 0 & 0 & 0 & 1
\end{bmatrix},
\end{equation*}
$T$ is the sampling period and $\pnoise_k \sim \Ndist(\pnoise_k;\vec{0},\vec{Q})$ is the process noise at time $k$, with
\begin{equation*}
\vec{Q} =
\begin{bmatrix}
q_1 \frac{T^3}{3} & q_1\frac{T^2}{2} & 0 & 0 & 0\\
q_1\frac{T^2}{2} & q_1T & 0 & 0 & 0\\
0 & 0 & q_1 \frac{T^3}{3} & q_1\frac{T^2}{2} & 0\\
0 & 0 & q_1\frac{T^2}{2} & q_1T & 0 \\
0 & 0 & 0 & 0 & q_2
\end{bmatrix},
\end{equation*}
where $q_1$ and $q_2$ are parameters of the model.

In order to isolate the benefits of iterating over the time update, a linear positional measurement model is used, i.e.,
\begin{equation}
\obs_k = \vec{H}\state_k + \onoise_k,
\end{equation}
with $\vec{H} = \mathrm{diag}\begin{bmatrix} 1 & 0 & 1 & 0 & 0 \end{bmatrix}$ and $\onoise_k\sim\Ndist(\onoise_k; \vec{0}, \sigma^2\vec{I})$.

The prior at time $0$ is given by
\begin{equation*}
p(\state_0) = \Ndist(\state_0; \hat{\state}_{0|0}, \vec{P}_{0|0}),
\end{equation*}
with $\hat{\state}_{0|0}^\top=\begin{bmatrix} 130 & 35 & -20 & -20 & -4\frac{\pi}{180} \end{bmatrix}$ and ${\vec{P}_{0|0}=\mathrm{diag}\begin{bmatrix} \sigma^2_{p_x} & \sigma^2_{v_x} & \sigma^2_{p_y} & \sigma^2_{v_y} & \sigma^2_{\omega}\end{bmatrix}}$, with $\sigma^2_{p_x}\!=\!\sigma^2_{v_x}\!=\!\sigma^2_{p_y}\!=\!\sigma^2_{v_y}\!=\!5$ and $\sigma^2_{\omega}\!=\!10^{-2}$.
The initial state for the ground truth trajectories are drawn from this prior.

\begin{figure}[tb]
\centering
\setlength{\textfloatsep}{1em}
\setlength{\floatsep}{1em}
\includegraphics[width=.8\columnwidth]{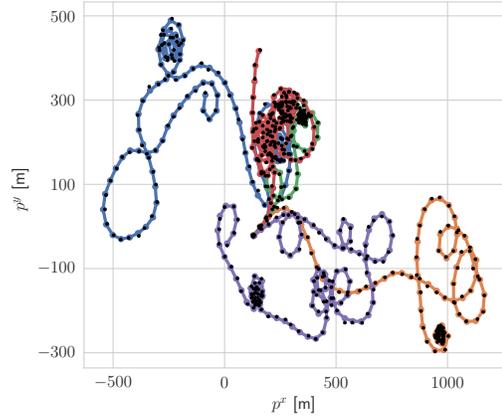}
\caption{Five example trajectories from one noise configuration of the considered tracking problem. Each trajectory is depicted as a separate color. The black smaller dots are a specific measurement realization along each trajectory.}
\label{fig:trackingexample}
\end{figure}

We fix $q_2=\num{e-2},T=1$ and sweep over all pairs of
\begin{align*}
q_1 &= \{\num{e-4},\num{e-3},\num{e-2},\num{e
-1},10^0\}\\
\sigma^2 &= \{\num{e-2},\num{e-1},10^0,\num{e+1},\num{e+2}\},
\end{align*}
i.e., 25 different noise configurations.
For each noise configuration, we simulate $10$ individual targets along $20$ different trajectories of length $K=130$ time steps, for a total of $200$ simulations per configuration.
Note that the $20$ trajectories are different for each noise configuration and that the $10$ targets for each trajectory differ only in their measurement noise realization.
However, the trajectories and measurement noise realizations are exactly the same for each algorithm.
Five example trajectories along with one measurement sample from each trajectory for a specific noise configuration is depicted in \cref{fig:trackingexample}.

To evaluate the performance of each dynamically iterated filter, we calculate the average position and velocity \abbrRMSE (separately) over the simulations for each of the filters and their corresponding baselines.
We also compute a ``relative'' \abbrRMSE, relative the non-iterated counterpart, i.e.,
\begin{equation}\label{eq:relativermse}
V = \frac{\textsc{RMSE}_{\textrm{iter}}}{\textsc{RMSE}_{\textrm{base}}},
\end{equation}
where clearly, $V\in[0,\infty]$ and lower is better.
A relative score of $0.9$ thus translates to a $10\%$ lower \abbrRMSE as compared to the baseline.
This yields a ``quick glance'' picture of the expected \abbrRMSE performance improvement in each particular noise configuration for each respective algorithm.
For the \abbrDIEKF the non-iterated baseline is the \abbrEKF whereas for both the \abbrDIUKF and \abbrDIPLF the baseline is the \abbrUKF.

\begin{figure*}[tb]
\centering
\subfloat[\label{fig:positionrmse}Position \abbrRMSE. Clearly, the analytically linearized \abbrDIEKF improves the most over it's baseline the \abbrEKF which diverges in 22/25 configurations. The \abbrDIUKF (middle) and \abbrDIPLF (right) share similar performance where the \abbrDIPLF is slightly better for some configurations.]{
\begin{tikzpicture}
\node[label={[yshift=-1cm, xshift=.4cm]\large Position \abbrRMSE $\left[\SI{}{\meter}\right]$}, label=below:{Relative \abbrRMSE Iterated/Baseline}] {\includegraphics[width=.97\textwidth]{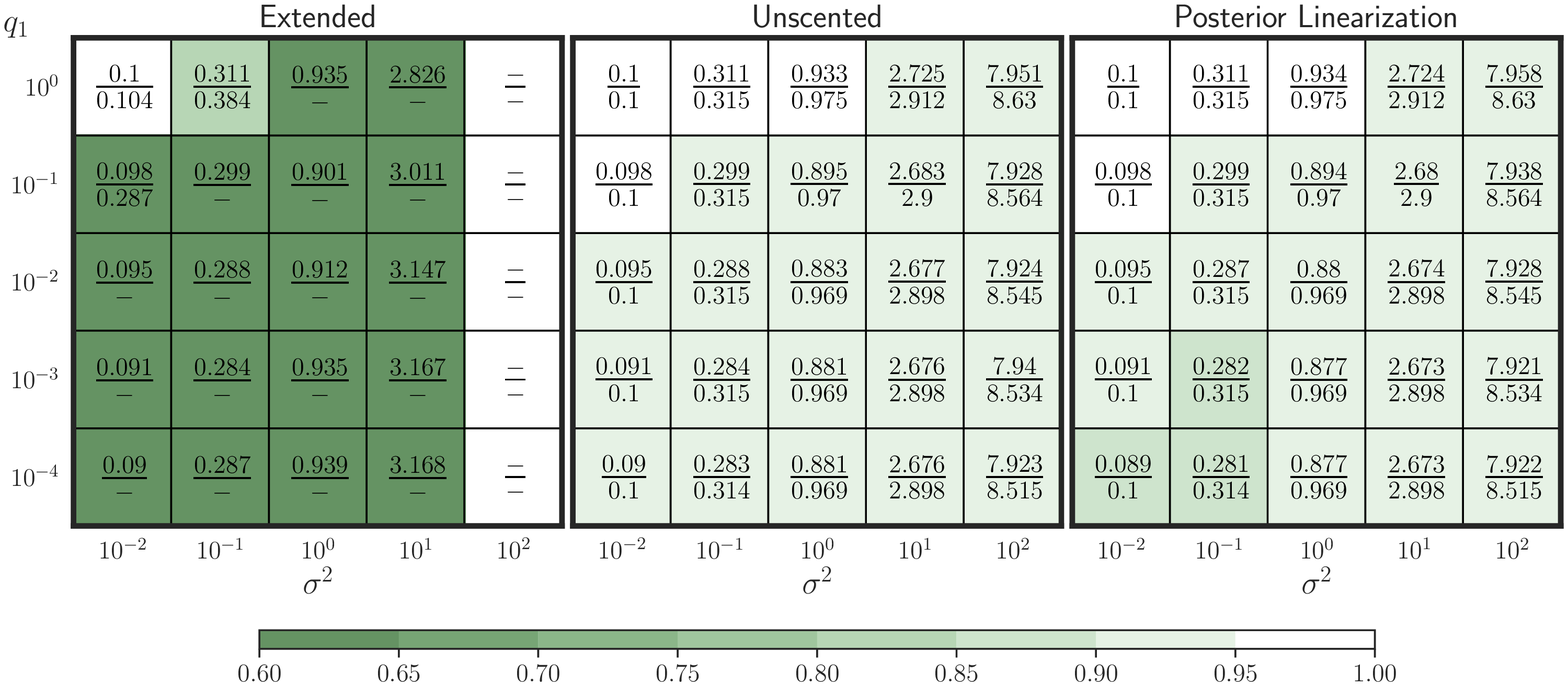}};
\end{tikzpicture}
}\\%
\subfloat[\label{fig:velocityrmse} Velocity \abbrRMSE. Clearly, all three dynamically iterated filters improve substantially over their corresponding baselines. For low noise regimes, up to 10-fold improvements are seen for the \abbrDIEKF whereas the \abbrDIUKF and \abbrDIPLF have approximately a 5-fold improvement. For high noise regimes the results for the \abbrDIUKF and \abbrDIPLF are even better with approximately 10-fold improvements.]{
\begin{tikzpicture}
\node[label={[yshift=-1cm, xshift=.35cm]\large Velocity \abbrRMSE $\left[\SI{}{\meter\per\second^{}}\right]$}, label=below:{Relative \abbrRMSE Iterated/Baseline}] {\includegraphics[width=.97\textwidth]{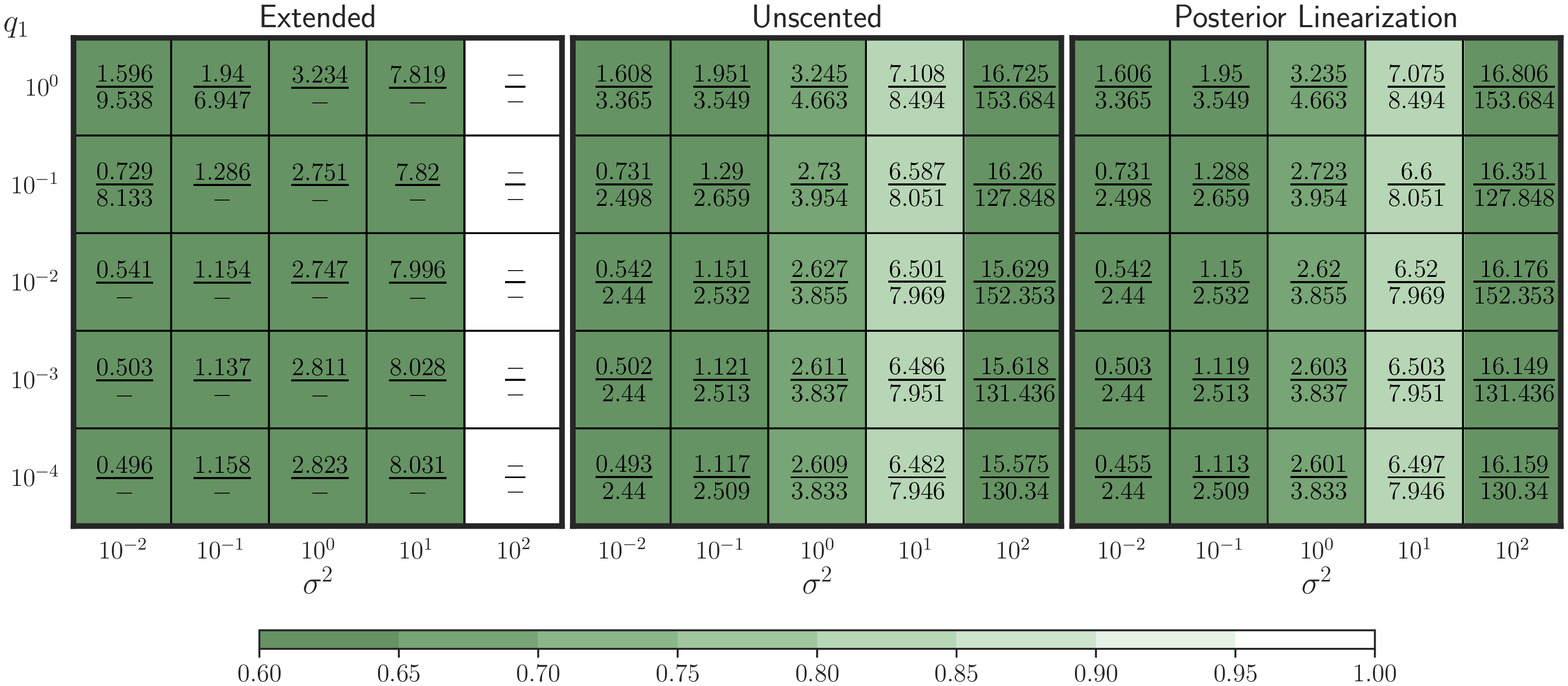}};
\end{tikzpicture}
}
\caption{\abbrRMSE for the dynamically iterated filters as compared to their respective baselines for 25 noise configurations, where each cell corresponds to one noise configuration given by the figure axes.
The top number in each cell is the \abbrRMSE for the dynamically iterated filter whereas the bottom number is the \abbrRMSE for the corresponding baseline.
A ``$-$'' indicates that the positional \abbrRMSE of the filter is larger than $\sigma$ and it is thus considered to have diverged, since an \abbrRMSE of $\sigma$ corresponds to just using the raw measurements.
The left plot in \protect\subref*{fig:positionrmse} and \protect\subref*{fig:velocityrmse} shows the \abbrRMSE for the \abbrDIEKF and \abbrEKF.
The middle figure shows the \abbrDIUKF and \abbrUKF and finally, the right most figure shows the \abbrDIPLF and \abbrUKF.
Each cell is colored according to the \abbrRMSE of the dynamically iterated filter relative to the baseline.
A relative \abbrRMSE of $0.9$ corresponds to a $10\%$ reduction of the \abbrRMSE.}
\label{fig:rmse}
\end{figure*}

The results are presented as $5\times 5$ matrices where each cell corresponds to a particular noise configuration for a particular pair of algorithms, e.g., the results for the \abbrDIEKF and \abbrEKF are summarized in one matrix.
The results can be found in \cref{fig:rmse} where the position and velocity {\abbrRMSE}s are presented in \cref{fig:positionrmse} and \cref{fig:velocityrmse}, respectively.
The leftmost matrix in each of the figures corresponds to the \abbrDIEKF and \abbrEKF. The middle matrix contains the results for the \abbrDIUKF and \abbrUKF and the rightmost matrix for the \abbrDIPLF and \abbrUKF.
The top number in each cell is the \abbrRMSE for the dynamically iterated filter whereas the bottom number corresponds to the baseline.
The color of each cell represents the \abbrRMSE of the dynamically iterated filter relative its baseline, according to \cref{eq:relativermse}.
A deeper green color thus indicates a more substantial improvement than a lighter green.
Lastly, an algorithm is considered to have diverged if its position \abbrRMSE is approximately larger than $\sigma$, where $\sigma$ is the measurement noise standard deviation, as a position \abbrRMSE of $\sigma$ can be expected by just using the raw measurements.
Divergence is illustrated by a ``$-$'' in the corresponding cell in the matrices.

%%%%%%%%%% RMSE over time figure
% \begin{figure}[tb]
% \centering
% \includegraphics[width=\columnwidth]{rmse-over-time}
% \caption{\abbrRMSE over time for the noise configuration $q_1=\num{e-3},\sigma^2=\num{e+2}$. The \abbrEKF is not visualized as it diverged for this configuration. The performance of the \abbrDIPLF and \abbrDIUKF is practically identical and improves over their baseline the \abbrUKF. The \abbrDIEKF is slightly worse but more or less matches the \abbrUKF for the velocity components. However, given that the baseline \abbrEKF diverged, the results for the \abbrDIEKF are still good.}
% \label{fig:rmseovertime}
% \end{figure}
%%%%%%%%%%%%

From \cref{fig:positionrmse}, it is clear that even though all of the dynamically iterated filters improve upon their baselines, the analytically linearized \abbrDIEKF benefits the most from the iterative procedure.
Astonishingly, the \abbrEKF diverges for 22 out of 25 configurations whereas the \abbrDIEKF manages to lower that to 5 out of 25 and only diverges in the high noise scenario ($\sigma^2=\num{e+2}$).
The performance increase in position \abbrRMSE is more modest for the \abbrDIUKF and \abbrDIPLF but still sees improvement, particularly for low process noise regimes.
For the velocity \abbrRMSE in \cref{fig:velocityrmse}, the improvement for all of the three dynamically iterated filters is substantial.
For low process noise regimes the improvement is up to 10-fold for the \abbrDIEKF and 5-fold for the \abbrDIUKF and \abbrDIPLF.
Even for modest noise levels, the \abbrDIUKF and \abbrDIPLF roughly manage a 2-fold performance improvement.
For the high noise scenario ($\sigma^2=\num{e+2}$), the \abbrDIUKF and \abbrDIPLF show a 10-fold performance improvement and bring the velocity \abbrRMSE down to reasonable levels where the \abbrRMSE for the \abbrUKF is very high.

% From \cref{fig:rmse}, we choose the noise configuration ${q_1=\num{e-3},\sigma^2=\num{e+2}}$ for additional evaluation, as this configuration has a reasonable measurement noise level and there is a clear difference between the performance of the \abbrDIEKF and the two other iterated algorithms.
% The \abbrRMSE in each individual state component over time are visualized for this noise configuration in \cref{fig:rmseovertime}.
% The \abbrRMSE is presented for the \abbrUKF, \abbrDIEKF, \abbrDIUKF and \abbrDIPLF but not the \abbrEKF as it diverged for this configuration and is hence left out for clarity.
%
% From \cref{fig:rmseovertime} it is clear that the \abbrDIUKF and \abbrDIPLF perform practically identically for this example and improves over their baseline the \abbrUKF.
% In an \abbrRMSE sense, the \abbrDIEKF is the worst. However, seeing as it's baseline, the \abbrEKF, diverges, the performance is still impressive.
% The \abbrRMSE in $\omega$ is slightly better for all iterated algorithms as compared to the \abbrUKF but is practically identical otherwise.

\section{Conclusion}\label{sec:conclusion}

Dynamically iterated filters, a new class of iterated nonlinear filters, has been presented.
The dynamically iterated filters, as opposed to previous iterated filters, are applicable to all possible combinations of (Gaussian) linear and nonlinear transition and measurement models.
The filters were evaluated against their respective non-iterated baselines in a numerical example with a nonlinear transition model and a linear measurement model.
Even in this ``simple'' case, where standard filters typically perform well, the dynamically iterated filters had improved \abbrRMSE performance, especially for non-measurable states.
Further, even though the \abbrEKF diverged in 22 out of 25 configurations considered, the dynamically iterated \abbrEKF was empirically shown to be stable for 20 out of 25 noise configurations, only diverging for high noise $(\sigma^2=\num{e+2})$.

Future work includes more extensive testing on other models as well as determining in what particular scenarios the statistically linearized versions perform better than the analytically and vice versa.

\section{Acknowledgments}\label{sec:acknowledgment}

The authors would like to extend sincere gratitude to Martin Skoglund for excellent tips for the experimental evaluation.

\appendices
\crefalias{section}{appendix}
\section{Loss Derivation}\label{app:loss}
The \abbrKL divergence between the true joint \cref{eq:jointdecomposition} and the approximate \cref{eq:approximatejoint} is given by

\vspace{1cm}

\begin{strip}
\scalebox{0.77}{\parbox{\linewidth}{%
\begin{align*}
&\mathrm{KL}\left(\fulljoint \Vert \approxjoint \right)=
\int \fulljoint\log \frac{\fulljoint}{\approxjoint}d\state_{k-1:k}d\auxil_{k-1:k}=\\
&\int \fulljointexpand
\log\frac{\fulljointexpand}{\approxjointexpand}d\state_{k-1:k}d\auxil_{k-1:k}=\\
&\int \fulljointexpand
\left[
\log\frac{\statejoint}{\approxstatejoint} +
\log\frac{\auxilcond{k}}{\approxauxilcond{k}} +
\log\frac{\auxilcond{k-1}}{\approxauxilcond{k-1}}
\right]d\state_{k-1:k}d\auxil_{k-1:k}=\\
&\int \fulljointexpand \log\frac{\statejoint}{\approxstatejoint}d\state_{k-1:k}d\auxil_{k-1:k} +
\int \fulljointexpand \log\frac{\auxilcond{k}}{\approxauxilcond{k}}d\state_{k-1:k}d\auxil_{k-1:k} + \\
&\int \fulljointexpand \log\frac{\auxilcond{k-1}}{\approxauxilcond{k-1}}d\state_{k-1:k}d\auxil_{k-1:k} =\\
&\underbrace{\int \statejoint \log\frac{\statejoint}{\approxstatejoint}d\state_{k-1:k}}_%
{\mathrm{KL}\left( \statejoint \Vert \approxstatejoint \right)} +
\underbrace{\int \statemarginal{k} \auxilcond{k} \log\frac{\auxilcond{k}}{\approxauxilcond{k}}d\state_kd\auxil_k}_
{\mathbb{E}_{\statemarginal{k}}\left[ \mathrm{KL}(\auxilcond{k}\Vert\approxauxilcond{k}) \right]} +
\underbrace{\int \statemarginal{k-1} \auxilcond{k-1} \log\frac{\auxilcond{k-1}}{\approxauxilcond{k-1}}}_%
{\mathbb{E}_{\statemarginal{k-1}} \left[ \mathrm{KL}(\auxilcond{k-1}\Vert\approxauxilcond{k-1}) \right]}d\state_{k-1}d\auxil_{k-1} = \\
&\mathrm{KL}\left( \statejoint \Vert \approxstatejoint \right) +
\mathbb{E}_{\statemarginal{k}}
\left[\mathrm{KL}(\auxilcond{k}\Vert\approxauxilcond{k})\right] +
\mathbb{E}_{\statemarginal{k-1}}
\left[\mathrm{KL}(\auxilcond{k-1}\Vert\approxauxilcond{k-1})\right] \triangleq \Loss.
\end{align*}}}
\end{strip}

\bibliographystyle{IEEEtran}
\bibliography{ms}
\end{document}